\documentclass{article}
\usepackage{spconf}
\usepackage{OIKF}
\usepackage{cite}

\usepackage[table, dvipsnames]{xcolor}
\usepackage{tikz}
\usetikzlibrary{positioning}

\usepackage{multirow}
\usepackage{subcaption}
\usepackage{makecell}
\usepackage{xcolor}
\usepackage{algorithm}
\usepackage{algpseudocode}
\usepackage{amsmath} %
\usepackage{amssymb}
\usepackage{tcolorbox}
\usepackage{romannum}
\usepackage[bookmarks,colorlinks]{hyperref}
\usepackage{cite}
\usepackage{array}
\newcolumntype{C}[1]{>{\centering\let\newline\\\arraybackslash\hspace{0pt}}m{#1}}
\newcolumntype{P}[1]{>{\centering\arraybackslash}p{#1}}
\newcommand{\figSpace}{ }
\newcommand{\EmMoment}{{\hat{\gvec{X}}_{t}^i}}

\title{Outlier-Insensitive Kalman Filtering using NUV Priors}
\name{Shunit Truzman,  Guy Revach, Nir Shlezinger, and Itzik Klein
\thanks{
		S. Truzman and I. Klein are with the Hatter Dept. of Marine Technologies, University of Haifa, Haifa, Israel, (e-mail: shunitruzman@gmail.com, kitzik@univ.haifa.ac.il ). 
		 G. Revach is with the Institute for Signal and Information Processing (ISI), D-ITET, ETH Zürich, 
		Switzerland (e-mail: grevach@ethz.ch). 
		N. Shlezinger is with the School of ECE, Ben-Gurion University of the Negev, Beer Sheva, Israel (e-mail: nirshl@bgu.ac.il). 
		S. Truzman. is supported by the Maurice Hatter Foundation.
		We thank Hans-Andrea Loeliger for helpful discussions.
	}}
\address{\vspace{-25mm}}
\begin{document}
\maketitle
%
%
\begin{abstract}
The \ac{kf} is a widely-used algorithm for tracking the latent state of a dynamical system from noisy observations. For systems that are well-described by \acl{lg} \acl{ss} models, the \ac{kf} minimizes the \ac{mse}. However,  in practice, observations are corrupted by outliers,  severely impairing the \ac{kf}'s performance. In this work, an outlier-insensitive \ac{kf} is proposed, where robustness is achieved by modeling each potential outlier as a \ac{nuv}. The \acp{nuv} variances are estimated online, using both \ac{em} and \ac{am}. The former was previously proposed for the task of smoothing with outliers and was adapted here to filtering, 
while both \ac{em} and \ac{am}  obtained the same performance and outperformed the other algorithms,  the AM approach is less complex and thus requires $40\%$ less runtime.
Our empirical study demonstrates that the \ac{mse} of our proposed outlier-insensitive \ac{kf} outperforms previously proposed algorithms, and that for data clean of outliers, it reverts to the classic \ac{kf}, i.e., \ac{mse} optimality is preserved.
\end{abstract}
%
%
\acresetall 
%
\vspace{-3mm}
\section{Introduction}\label{sec:intro}
\vspace{-2mm}
\acl{Se} of dynamical systems from noisy observations plays a key role in various scientific and technological fields such as radar target tracking, complex image processing,  navigation, and positioning~\cite{Koopman2012}. 
The celebrated \ac{kf}~\cite{R.E.Kalman1960} is an efficient recursive state estimation algorithm that is  
\ac{mse} optimal for dynamical systems obeying a \acl{lg} \ac{ss} model. However, the quadratic form of its objective, i.e., \ac{mse}, makes it sensitive to deviations from nominal noise. Thus, the \ac{kf} is severely impaired when outliers are present in the measurements~\cite{Aravkin2017,Farahmand2011}.  As sensory data is often populated with outliers,  robustness to outliers is essential  \cite{Ting2007,Agamennoni2011,Wadehn2016}.
A common approach for dealing with outliers is to detect and then disregard influential observations. Such detection can be achieved using appropriate statistical diagnostics~\cite{alma99117188325005503} on the posterior distribution, e.g., $\ztest$~\cite{Rousseeuw2011} and $\chitest$~\cite{Ye2001, VanWyk2020}.  
The main drawbacks of these approaches are that they need to be carefully tuned for a required false alarm, and that potentially useful outlier information is not accounted for in the estimation process. Alternatively, one can limit the effect of outliers by reweighting the covariance of the observation noise at each data sample when estimating the current state~\cite{Ting2007, Agamennoni2011}. These techniques require careful tuning of multiple hyperparameters to operate reliably as well. A different approach formulates the \ac{kf} as a linear regression problem, detecting outliers via a sparsifying $\ell_1$-penalty \cite{Farahmand2011, Aravkin2011}, tackled via optimization techniques, which may be computationally complex. 

In this work, a new approach for \ac{oikf} is proposed that leverages ideas from \emph{sparse} Bayesian learning \cite{Wipf2011a}. Here, each potential outlier is modeled as a \ac{nuv} \cite{Loeliger2017, Loeliger2019, Wadehn2016}, i.e., an additive component on top of the observation noise. \ac{nuv} incorporation effectively yields a modified overall sparsity-aware objective~\cite{Loeliger2017}, which is shown to yield a robust outlier detection statistical test with a relatively low false alarm. 
When an outlier is reliably detected, we incorporate its variance into the overall covariance matrix of the observation noise, thereby balancing its contribution to the information fusion (i.e., update) step in the \ac{kf}, and the outlier information is used in the \acl{se} process.

To estimate the \ac{nuv} online, we first adapt the \ac{em} algorithm~\cite{Hutchinson1979,Stoffer1982,SophoclesJ.Orfanidis2018}, which was previously proposed for offline smoothing~\cite{Wadehn2016}, to the online filtering task. \ac{em} is based on computing second-order moments, i.e., the full state observation posterior covariance. We then present an implementation, which has not been considered to date, using a \emph{simpler} \ac{am} algorithm~\cite{AAndresen2016,M.I.JordanZ.GhahramaniT.S.JaakkolaandL.K.Saul1999,F.BachR.JenattonJ.Mairal2012}. Unlike \ac{em}, \ac{am} uses only first-order moments as an empirical surrogate for outlier detection, without sacrificing performance and with improved robustness. 
We evaluate the \ac{oikf} for tracking based on the \ac{wna} motion model with outliers~\cite{bar2004estimation}.
We empirically demonstrate the superiority of our proposed algorithm compared to previous robust variants of the \ac{kf}, achieving improved performances with low complexity operations.

%
%
The rest of this paper is organized as follows: \secref{sec:SysModel} formulates the system and the \ac{kf} with outliers. \secref{sec:OIKF} presents the \ac{oikf} with its estimation for the outlier's variance, and \secref{sec:NumEval} empirically evaluates our methods.
%
%
\section{System Model and the Kalman Filter}\label{sec:SysModel}
%
\ac{ss} models in \acl{dt} are a common characterization of dynamical systems \cite{bar2004estimation}. Such representations capture the relationship between an unknown latent state vector $\gvec{x}_t$ and an observed vector $\gvec{y}_t$, where $t\in\gint$ is the time index. Here, a Gaussian and continuous \ac{ss} model is considered, namely
\begin{subequations}
\label{eq:ssmodel}
\begin{align}
\label{eq:ssmodelx}
\gvec{x}_{t}&= 
\gvec{F}\cdot{\gvec{x}_{t-1}}+\gvec{e}_{t},
\quad
\,\,\,\,\,\,\gvec{e}_t\sim 
\mathcal{N}\brackets{\gvec{0},\gvec{Q}},
\quad
\gvec{x}_{t}\in\greal^m\\
\label{eq:ssmodely}
\gvec{y}_{t}&=
\gvec{H}\cdot\gvec{x}_t+\gvec{z}_t+\gvec{u}_t,
\,\,\,\,\gvec{z}_t\sim
\mathcal{N}\brackets{\gvec{0},\gvec{R}},
\quad
\gvec{y}_{t}\in\greal^n.
\end{align}
\end{subequations}

In \eqref{eq:ssmodelx}, the state evolves by an evolution matrix $\gvec{F}$ and by \ac{awgn} $\gvec{e}_{t}$ with covariance matrix $\gvec{Q}$. In \eqref{eq:ssmodely}, the state observations are generated by the linear mapping $\gvec{H}$ corrupted by an \ac{awgn} $\gvec{z}_{t}$ with a diagonal covariance matrix $\gvec{R} = \textrm{diag}\brackets{\gvec{r}^2}$, and by additive outlier impulsive noise $\gvec{u}_{t}$ with an \emph{unknown} distribution.
%
%
The \ac{kf} is an efficient online recursive filter that estimates the state $\gvec{x}_t$ from the observations $\{\gvec{y}_\tau\}_{\tau\leq t}$. It is \ac{mse} optimal for the \ac{ss} model \eqref{eq:ssmodel} \emph{without} the outliers $\gvec{u}_t$. It can be conceptualized as a two-step procedure in each time step $t$, \textit{predict} and \textit{update}, in which the joint probability  distribution over the variables is computed, using the first- and second-order moments of the Gaussian distribution. In the \emph{predict} step, the \emph{prior} distribution is computed, namely
%
%
\begin{subequations}
\begin{align}
\label{eqn:predict}
\hat{\gvec{x}}_{t\given{t-1}} &= 
\gvec{F}\cdot{\hat{\gvec{x}}_{t-1}},
\hspace{0.45cm}
\mySigma_{t\given{t-1}} \!=\!
\gvec{F}\cdot\mySigma_{t-1}\cdot\gvec{F}^\top\!+\!\gvec{Q},\\
\label{eqn:predict2}
\hat{\gvec{y}}_{t\given{t-1}} &=
\gvec{H}\cdot{\hat{\gvec{x}}_{t\given{t-1}}},
\hspace{0.2cm}
\gvec{S}_{t\given{t-1}}\! =\!
\gvec{H}\cdot\mySigma_{t\given{t-1}}\cdot\gvec{H}^\top\!+\!\gvec{R}.
\end{align}
\end{subequations}

In the \emph{update} step the \emph{posterior} distribution is computed by fusing the new observation $\gvec{y}_t$ with the previously predicted prior $\hat{\gvec{x}}_{t\given{t-1}}$, where the \ac{kg} matrix $\Kgain_t$ is used to \emph{balance} the contributions of both parts, namely
%
%
\begin{equation}
\hat{\gvec{x}}_{t} = 
\hat{\gvec{x}}_{t\given{t-1}}+\Kgain_{t}\cdot\Delta\gvec{y}_t,\hspace{-0.24cm}
\quad
{\mySigma}_{t}\! =\!
{\mySigma}_{t\given{t-1}}\!-\!\Kgain_{t}\cdot{\mathbf{S}}_{t\given{t-1}}\cdot\Kgain^{\top}_{t},
\label{eqn:update}
\end{equation} 
%
%
\begin{equation}\label{eq:kgain}
\Kgain_{t}={\mySigma}_{t\given{t-1}}\cdot{\gvec{H}^\top}\cdot{\gvec{S}}^{-1}_{t\given{t-1}}, 
\quad
\Delta\gvec{y}_t=\gvec{y}_t-\hat{\gvec{y}}_{t\given{t-1}}.
\end{equation}
%
\section{OUTLIER-INSENSITIVE KALMAN filtering}\label{sec:OIKF}
%
Next we propose our \emph{outlier-robust} online filter. In \ssecref{sec:nuv} we present the \ac{nuv} modeling. Then we present the \ac{oikf} algorithm in \ssecref{subsec:OIKFAlg}, after which we derive two considered methods for \ac{nuv} estimation based on \ac{em} (\ssecref{subsec:EM}) and \ac{am} (\ssecref{subsec:AM}).
%
%
\begin{figure}[t]
\centering
\vspace{-3mm}
\includegraphics[width=8cm,height=4.5cm]{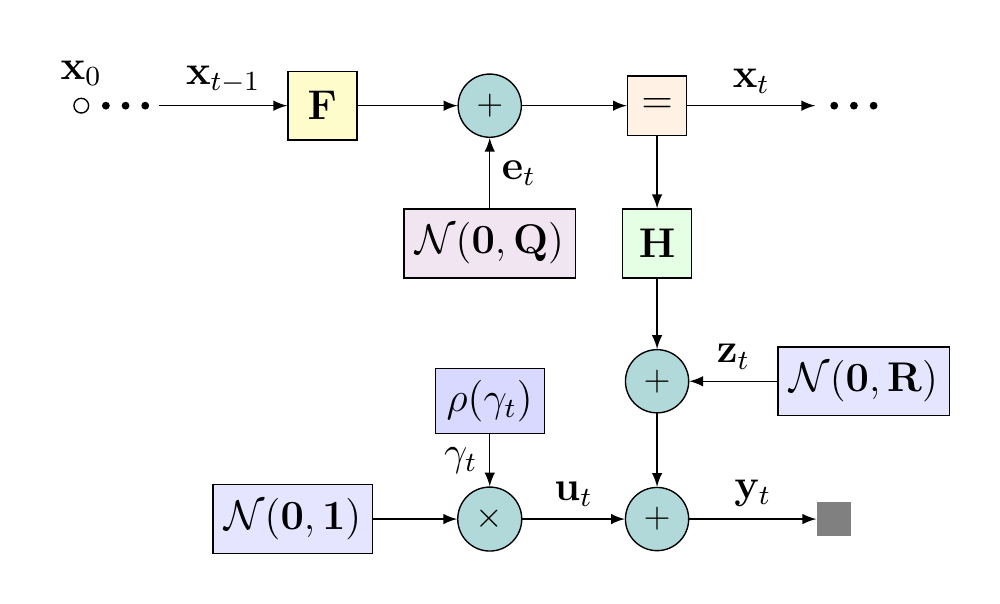}
\vspace{-4mm}
\caption{Factor graph of the system model at time step $t$}\label{fig:SSmodel}
\end{figure}
%
%
\vspace{-2mm}
\subsection{NUV Modeling}\label{sec:nuv}
\vspace{-1mm}
Given an observation sample $\gvec{y}_{t}$ 
\eqref{eq:ssmodely}, we define $\gvec{v}_t$ to be the error vector as the sum of two independent sources: the observation noise $\gvec{z}_t$ and the outlier-causing impulsive noise $\gvec{u}_t$, modeled as \ac{nuv} with unknown variance vector ${\vgamma}^2_t$. Thus, the covariance of  $\gvec{v}_t$, $\gvec{\Gamma}_t$, is diagonal and comprises the sum of variances of the two noise sources, namely
\begin{subequations}
\begin{align}
\label{eqn:v}
\gvec{v}_t&\triangleq\gvec{y}_t-\gvec{H}\cdot\gvec{x}_t=\gvec{z}_t+\gvec{u}_t,
\quad
\gvec{u}_t\sim\mathcal{N}\brackets{\gvec{0},{\vgamma}^2_t},
\\
\label{eqn:R}
\gvec{v}_t&\sim\mathcal{N}\brackets{\gvec{0},\gvec{\Gamma}_t},
\quad
\gvec{\Gamma}_t=
\textrm{diag}\brackets{{\vnu}^2_t},
\quad
{\vnu}^2_t\triangleq\gvec{r}^2+{\vgamma}^2_t.
\end{align}
\end{subequations}
The motivation for utilizing the \ac{nuv} framework stems from its ability to systematically incorporate interference of a bursty nature~\cite{Loeliger2017, Loeliger2019}, and is thus useful for handling outliers with sparse \acl{ls} (quadratic) models~\cite{Wadehn2016}. 
The incorporation of the \ac{nuv} representation to the overall \ac{ss} model \eqref{eq:ssmodel}, is illustrated as a factor graph~\cite{Loeliger2007} in \figref{fig:SSmodel}.
%
%
\vspace{-2mm}
\subsection{OIKF Algorithm}\label{subsec:OIKFAlg}
The proposed \ac{oikf} uses \ac{map} estimation to instantaneously estimate the unknown variance ${\vgamma}^2_t$. By either applying \ac{em} or \ac{am} in each time step $t$, we obtain an estimate for the variance. In the \ac{em} version, the  second-order moment $\gvec{\nu}_t$~\eqref{eqn:R} is directly estimated, while in \ac{am} version it is obtained from estimating $\gvec{v}_t$~\eqref{eqn:v}, i.e., from the first-order moment. 
The \ac{oikf} thus uses
\begin{equation}\label{eq:nuv_est}
{\hat{\vgamma}}^2_t=\max\set{\vnu_t^2-\gvec{r}^2,0},
\hspace{0.175cm}
\vnu_t^2=\set{\textrm{EM}:\hat{\vnu}_t^2,\textrm{AM}:\gvec{\hat{v}}^2_t}.
\end{equation}

A key property of the \ac{map} estimator \eqref{eq:nuv_est}
is that it tends to be \emph{sparse} \cite{Loeliger2017}, thus providing a \emph{robust} statistical test to detect the presence of outliers. Due to the sparsity property, in most of the time steps, outliers are not detected, and filtering coincides with the standard \ac{kf}, thus preserving its optimality for data without outliers. When outlier is detected, namely, when $\hat{\vgamma}^2_t\neq\gvec{0}$, its contribution to the update step is balanced via its estimated variance; it is integrated into the overall error covariance $\gvec{\Gamma}_t$, which in turn becomes the observation noise and therefore affects the \ac{kg} in the update equation. 

The above procedure is repeated iteratively, and the pseudo-code for the proposed  \ac{oikf} is summarized in Algorithm~\ref{alg:alg1}. It can be done for  fixed $N$  iterations, or until convergence. In \cite{Wadehn2016}, $\hat{\vgamma}^2_t$ was initialized to $\gvec{0}$. Here, empirical experience suggests to initialize it to the value of the prior moments. While retaining performance, it converges faster.
%
%
\setlength{\textfloatsep}{3pt}
\begin{algorithm}[tb]
\begin{small}
\caption{\ac{oikf} at time instance $t$}\label{alg:alg1}
\begin{algorithmic}
\State Number of iterations $N$
\State \textbf{Predict:} Estimate \emph{a priori} for $\gvec{\hat x}_{t\given{t-1}
}^{i=0}, \gvec{ \mySigma}_{t\given{t-1}}^{i=0}$ via~\eqref{eqn:predict}
\For{$i = 0,...,N-1$}
     \State
\textbf{AM:} Compute $\gvec{\hat v}_{t}^i=\gvec{y}_t-\gvec{H}\cdot\gvec{\hat x}_{t}^i$ \\
\hspace{1.3cm}Estimate $\brackets{\hat\vgamma^{i}_{t}}^2$ via~\eqref{eqn:gammaAM} 
\State
\textbf{EM:} Estimate $\brackets{\hat\vgamma^{i}_{t}}^2$ via~\eqref{eq:gammaEM} with the $2$nd-order  \\\hspace{1.3cm} moment $\EmMoment$ as in~\eqref{eqn:momentII}
\State Compute $\gvec{\Gamma}_{t}^i=\textnormal{diag}\brackets{\gvec{r}^2+\brackets{\hat\vgamma^i_{t}}^2}$ 
 \State Compute $ \gvec{\hat y}_{t\given{t-1}}^i,\gvec{S}_{t\given{t-1}}^i$ via ~\eqref{eqn:predict2}  with $\gvec{R}=\gvec{\Gamma}_{t}^i$.
\State \textbf{Update:} Estimate \emph{a posteriori} for $\gvec{\hat x}_{t}^i, \mySigma_{t}^i$ via~\eqref{eqn:update} 
\EndFor
\end{algorithmic}
\end{small}
\end{algorithm}
%
%
\vspace{-2mm}
\subsection{Expectation Maximization}
\label{subsec:EM}
For an observation sample $\gvec{y}_t$,
the \ac{map} estimate for $\hat{\vgamma}^2_t$ is
\begin{align}\label{eqn:EM}
{\hat\vgamma_{t}^2}\brackets{\gvec{y}_t} =
{\arg}\max_{\vgamma^2_t} p\brackets{\gvec{y}_t\given{\vgamma^2_t}}\cdot p\brackets{\vgamma^2_t}.
\end{align}
To compute the \ac{map} estimate using EM, we assume a \emph{plain} \ac{nuv}, i.e., a uniform prior $p\brackets{\vgamma^2_t}\propto1$~\cite{Loeliger2017}. We can now evaluate the standard \ac{em} by alternating between the E-step, i.e., the conditional expectation, and the M-step, i.e., maximizing this expression with respect to $\vgamma^2_t$. Here, the expectation step corresponds to the \ac{kf}, from which we get first- and second-order \emph{posterior} moments, namely
\vspace{-0.9mm}
\begin{equation}\label{eqn:momentII}
\hat{\gvec{x}}_{t}^i,
\qquad
\EmMoment\triangleq \mySigma^i_{t}+\hat{\gvec{x}}_{t}^i\cdot{\hat{\gvec{x}}^i_{t}}^\top.
\vspace{-0.9mm}
\end{equation}
The \emph{maximization} step  corresponds to recovering the covariance ~\eqref{eqn:R} using the following estimate~\cite{Wadehn2016,SophoclesJ.Orfanidis2018}:
\begin{equation}
{\gvec{\hat{\Gamma}}_{t}^i}=\gvec{y}_{t}\cdot\gvec{y}^\top_{t}-\gvec{H}\cdot\hat{\gvec{x}}_{t}^i\cdot{\gvec{y}^\top_{t}}-{\gvec{y}}_{t}\cdot{{\gvec{\hat{x}}^i_{t}}}^\top\cdot\gvec{H}^\top-\gvec{H}\cdot\EmMoment\cdot\gvec{H}^\top.
\label{nu}
\end{equation}

We exploit the fact that $\gvec{{\Gamma}}_{t}$ in \eqref{eqn:R} is diagonal, to estimate the variance $\brackets{\gscal{\hat{\nu}}^i_{t,k}}^2$ for each dimension $ k\in\{1,\dots,n\}$ in a scalar manner. Since ${\gamma}_{t,k}^2$ is non-negative, we get the following expression for its estimate:
\begin{equation}
\brackets{{\hat\gamma}^{i}_{t,k}}^2=\max\set{\brackets{{\hat\nu}^{i}_{t,k}}^2-\gscal{r}_k^2,0}.
\label{eq:gammaEM}
\end{equation} 
This demonstrates the sparsifying property of the \ac{nuv} modeling, maybe more specifically the \ac{map} estimate for $\gscal{u}_{t,k}$ from $\hat{\gamma}^2_{t,k}$ \cite{Loeliger2017} is 
\begin{align}
\hat{\gscal{u}}_{t,k}&= \gscal{v}_{t,k}\cdot\frac{\hat{\gamma}^2_{t,k}}{\hat{\gamma}^2_{t,k}+{\gscal{r}^2_{k}}}
=\max\set{\gscal{v}_{t,k}\cdot\frac{\hat{\nu}^2_{t,k}-{\gscal{r}^2_{k}}}{\hat{\nu}^2_{t,k}},0}.
\end{align}
Namely, $\hat{\gamma}^2_{t,k}$ leads to $\hat{\gscal{u}}_{t,k}=0$. When the outlier is identified as zero, \ac{oikf} coincides with the \ac{kf}.  
%
%
\vspace{-2mm}
\subsection{Alternating Maximization}\label{subsec:AM}
\vspace{-1mm}
We next describe an iterative \ac{am} method to compute the joint \ac{map} estimate based on the \ac{nuv} representation~\cite{Loeliger2019}
\begin{equation}
[\hat{\gvec{v}}_{t},\hat{\gscal{\vgamma}}^2_{t}](\gvec{y}_{t})
\!=\!\underset{{\gvec{v}_{t},\vgamma^2_{t}}}{\arg\max}\,p\brackets{{\gvec{y}_{t}\given{\gvec{v}_{t}}}}\cdot p\big(\gvec{v}_{t}\big|{\vgamma^2_{t}}\big)\cdot p\brackets{\vgamma^2_{t}}.
\end{equation}
\ac{am} iterates between a maximization step over the error state ${\gvec{v}}_{t}$ with fixed variance ${\vgamma^2_{t}}$:
\begin{equation}
\hat{\gvec{v}}_{t}\!=\!\underset{{\gvec{v}_{t}}}{\arg\max}\, p\brackets{{\gvec{y}_{t}\given{\gvec{v}_{t}}}}\cdot p\brackets{\gvec{v}_{t}\given\vgamma^2_{t}}\cdot p\brackets{\vgamma^2_{t}},
\end{equation}
and a maximization step over the unknown variance ${\vgamma^2_{t}}$ based on $\gvec{v}_{t}$, i.e., finding
\begin{equation}\label{eqn:AMMap}
 \hat{\vgamma}^2_{t}=
\underset{{\vgamma^2_{t}}\geq 0}{\arg\max}\,p\brackets{\gvec{v}_{t}\given \vgamma^2_{t}}\cdot p\brackets{\vgamma^2_{t}}.
\vspace{-3mm}
\end{equation}
Similarly to our derivation of  \ac{em} in Subsection~\ref{subsec:EM}, we assume a uniform prior on ${\vgamma}_{t}^2$. However, while \ac{em} utilizes estimates of both the first- and second-order moments of the states, namely, $\hat{\gvec{x}}_{t}^i$ and $\EmMoment$ \eqref{eqn:momentII}, \ac{am} uses only $\hat{\gvec{x}}_{t}^i$.

For convenience, we formulate the second step in a scalar manner, which extends to multivariate observations by assuming that the observation
noise $\gvec{z}_t$ and the outlier $\gvec{u}_t$ in each dimension are independent. In particular, by replacing $\gvec{v}_{t}$ in \eqref{eqn:AMMap} with its instantaneous estimate $\gvec{\hat v}_{t}^i=\gvec{y}_t-\gvec{H}\cdot\gvec{\hat x}_{t}^i$ computed from the \ac{kf}, we obtain the following update rule for the  $k$th entry of the unknown variance
\begin{equation}
\brackets{{\hat\gamma}^{i}_{t,k}}^2
\!=\!\max\set{{({\gscal{\hat v}^i_{t,k}})^2\!-\!{\gscal{r}^2_{k}},0}}.
\label{eqn:gammaAM}
\end{equation}
We obtain an analytic expression of $\hat\gamma^2_{t,k}$ for the update step, which is parameter-free. Furthermore, combining \ac{am} with the \ac{nuv}-prior results in an equivalent cost function  $\loss{\gvec{v}_{t}}\triangleq-\log\,p\brackets{\gvec{v}_{t}}$ that is non-convex~\cite{Loeliger2019}. This is equivalent loss function useful for sparse \acl{ls} models, e.g., \ac{kf} with outliers \cite{Loeliger2019}, and $\hat\vgamma_k^2$ is likely to be sparse, following the same arguments as in Subsection~\ref{subsec:EM}.

\begin{figure*}
\begin{center}
\begin{subfigure}[h]{0.65\columnwidth}
\includegraphics[width=5.5cm,height=3.8cm]{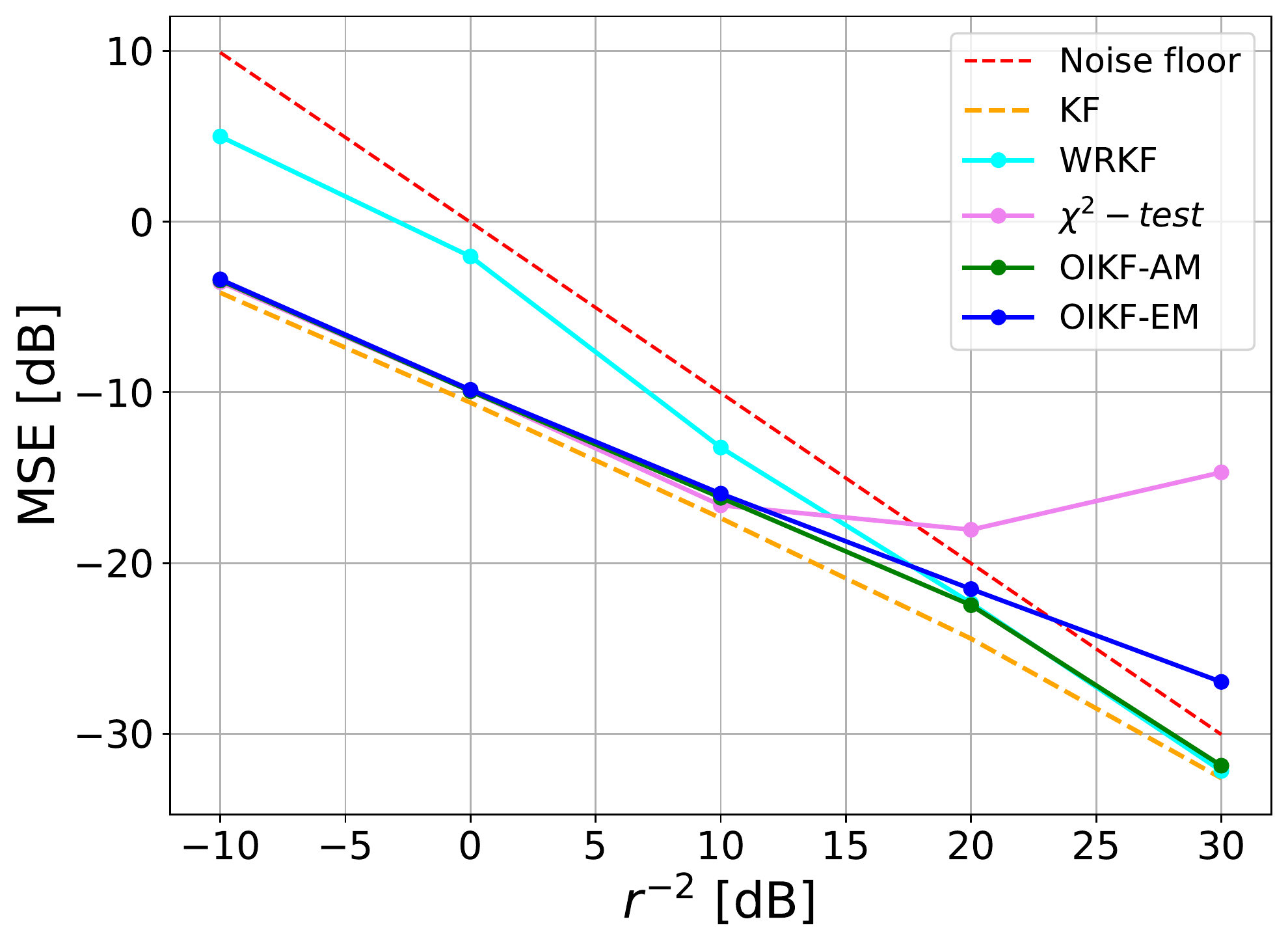}
\vspace{-1mm}
\caption{Noisy data clean of outliers}
\label{fig:MSE noisy}
\end{subfigure}
\figSpace
\begin{subfigure}[h]{0.65\columnwidth}
\includegraphics[width=5.5cm,height=3.8cm]{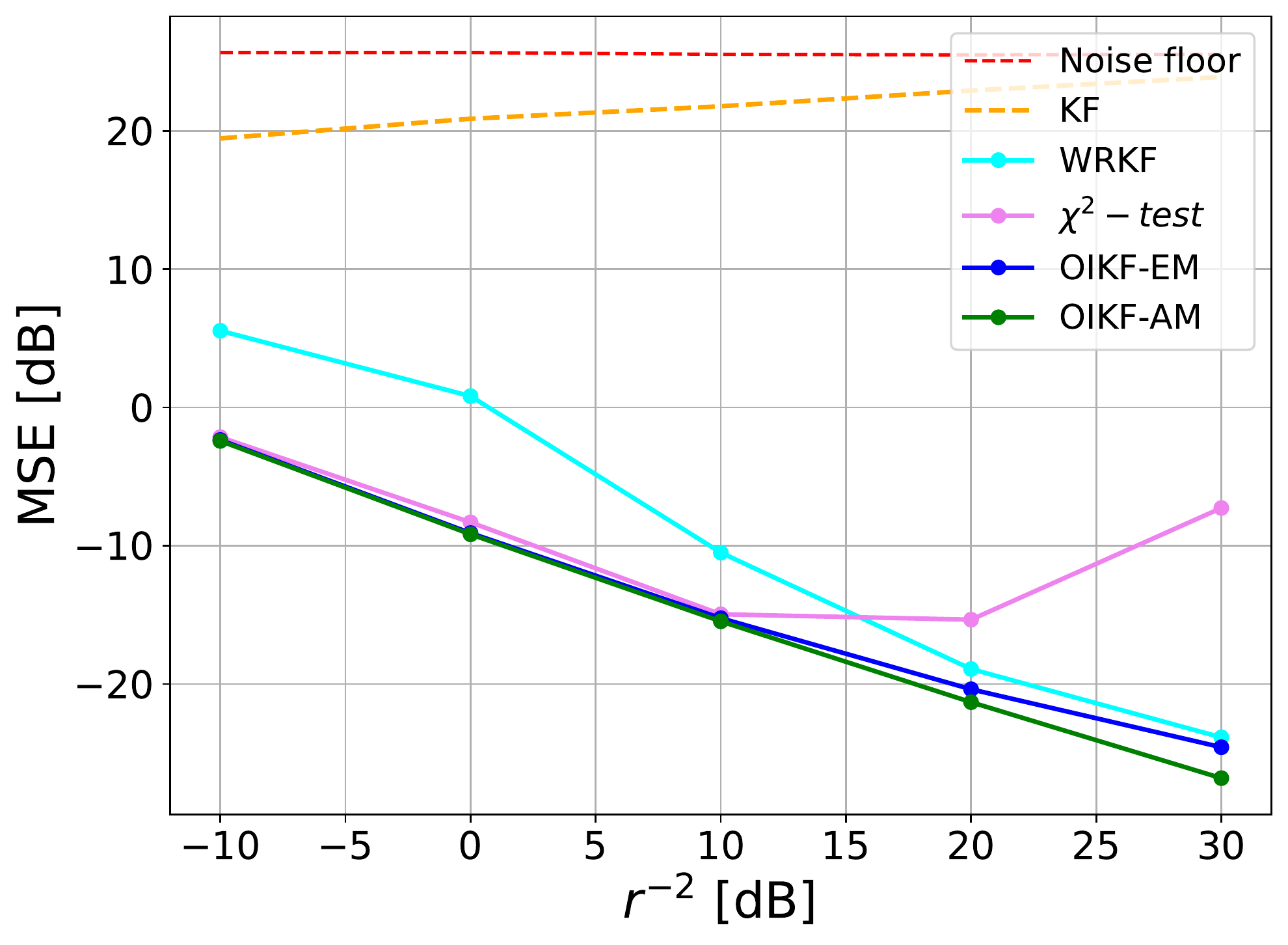}
\vspace{-1mm}
\caption{Noisy data with outliers}
\label{fig:MSE noisy with outliers}
\end{subfigure}
\figSpace
\begin{subfigure}[h]{0.65\columnwidth}
\includegraphics[width=5.5cm,height=3.8cm]{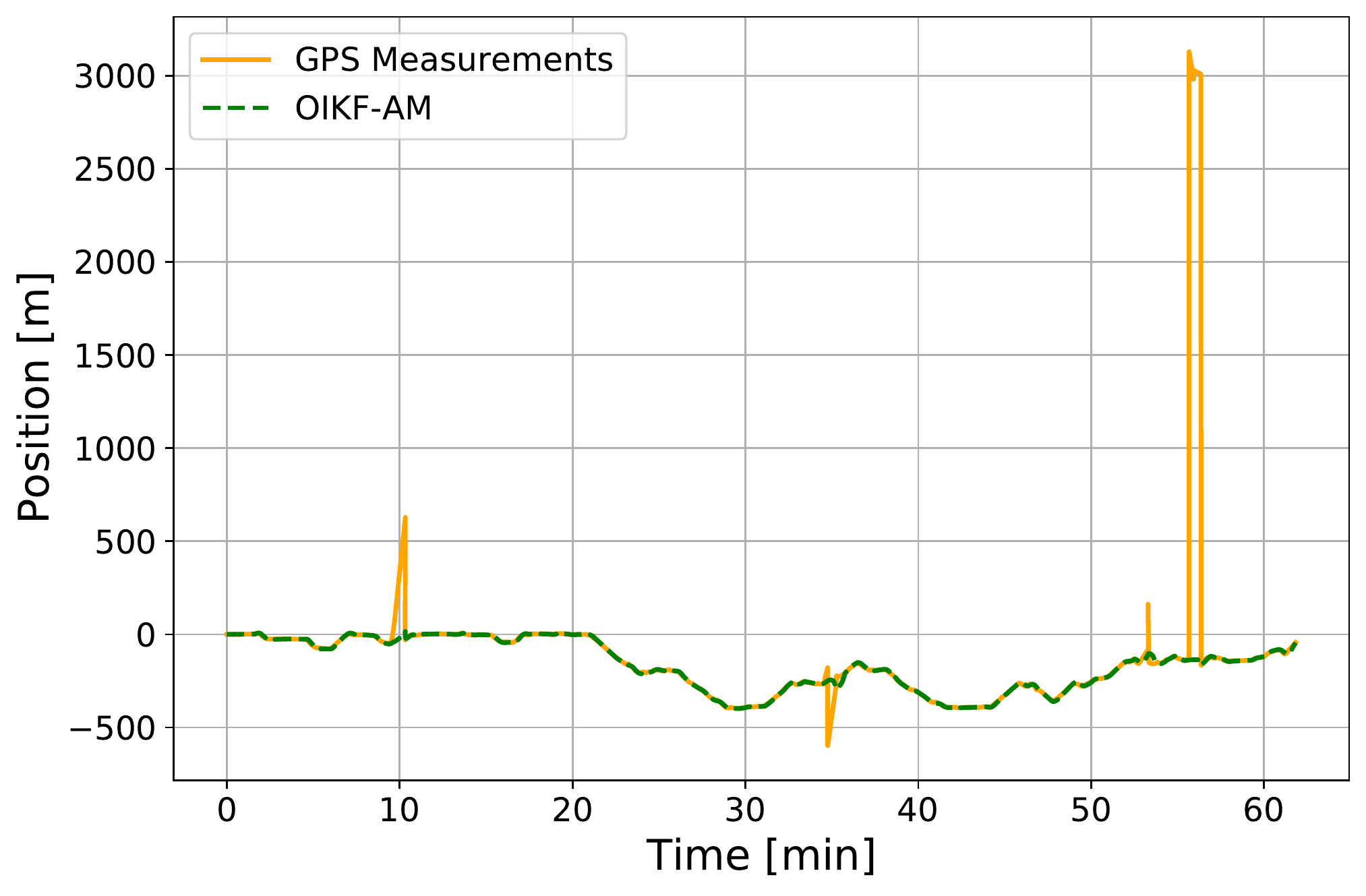}
\vspace{-1mm}
\caption{Vehicle estimated trajectory}
\label{fig:position estimation}
\end{subfigure}
\vspace{-1mm}
\caption{Sub-Fig.~\ref{fig:MSE noisy} and~\ref{fig:MSE noisy with outliers} present the obtained MSE for different algorithms and Sub-Fig.~\ref{fig:position estimation} presents the measured vehicle position obtained from the noisy GPS readings (orange line), and the trajectory estimated by our \ac{oikf}-\ac{am} (green dashed line).}
\end{center}
\figSpace
\vspace{-5mm}
\end{figure*}
%
%
\vspace{-2mm}
\subsection{Discussion}
\vspace{-1mm}
The proposed \ac{oikf} is designed to track in the presence of outliers by iteratively refining its update step with \ac{map} estimation of the outlier variance. 
Both \ac{em} and \ac{am} algorithms can be used to tackle this \ac{map} estimation, which arises from the \ac{nuv} modelling. The main difference between the algorithms is that \ac{em} uses the second-order moment of the state vector, while \ac{am} uses the first-order moment only, hence it does not require the posterior covariance matrix $\mySigma_t$ in each iteration. Consequently, \ac{am} avoids possibly unnecessary heuristics in the model noise, making it more robust compared to \ac{em}, as empirically demonstrated in Section~\ref{sec:NumEval}. The fact that \ac{am} does not explicitly rely on the second-order moments is expected to facilitate its augmentation with trainable data-driven variants of the Kalman filter, e.g.,~\cite{Revach2022,ni2022rtsnet,klein2022uncertainty}. Such a fusion of \ac{oikf} with data-aided computations, which we intend to explore in future work, bears the potential of facilitating robust filtering in partially known \ac{ss} models.

\ac{oikf} provides a degree of freedom in choosing the prior of  $\vgamma_k^2$.  We chose it to be uniform, which is parameter free, and can be shown to effectively modify the overall loss function to account for sparse outliers~\cite{Loeliger2019}.
Alternative settings of this prior would result in different effective loss functions such as the convex Huber cost function~\cite{Roncetti2009,Loeliger2019}. We leave the investigation of \ac{oikf} with different priors  to future work.
%
%
\vspace{-2mm}
\section{Empirical Evaluation}\label{sec:NumEval}
\vspace{-2mm}
To evaluate the proposed \ac{oikf}-\ac{am} and \ac{oikf}-\ac{em}, we tested these algorithms on a standard localization task\footnote{The source code and additional information on the empirical study can be found  at \url{https://github.com/KalmanNet/OIKF_ICASSP23}.}. 
We compare our proposed algorithms performance in terms of position error to  the following alternative algorithms: $1)$ the standard \ac{kf}; $2)$ the well-known $\chitest$ \cite{Ye2001} (with a $95\%$ confidence level); $3)$ the
weighted covariance methods, i.e., the WRKF from \cite{Ting2007}; and $4)$ the ORKF from \cite{Agamennoni2011}. 
We evaluate these algorithms in two settings: 
A synthetic dataset generated from a known SS model, and real-world dynamics data based on the Michigan NCLT dataset \cite{Carlevaris-Bianco2016}. In both cases,
the state vector is given by 
$\gvec{x}=\brackets{\gscal{p,v}}^\top$,
where $\gscal{p}$ and $\gscal{v}$ are the position and velocity, respectively. 

For the \textbf{synthetic data} we define the dynamic \ac{wna} model \cite{bar2004estimation}, defined by  a linear \ac{ss} model. The system is fully observable, thus the observation matrix is the identity matrix $\gvec{H}= \gvec{I}_n$ and the observation noise covariance matrix is defined as $\gvec{R}=\gscal{r}^2\cdot \gvec{I}_n$. The state-evolution noise variance $\textnormal{q}^2$ 
is set to a constant value of $-10\dB$. Uncertainty in the dynamics and measurements is accounted for by Gaussian i.i.d input, state, and measurements noise, whereas the outliers are modeled with intensity sampled from a Rayleigh distribution  $\mathcal{RAY}\brackets{\beta}$ 
with scale parameter $\beta=30$.
The outliers' time steps are drawn from a Bernoulli distribution, namely
$\mathcal{B}\brackets{p}$, where ${p}$ is set to 0.2.
For the \textbf{real-world data} we use the NCLT dataset from session with date 2013-04-05, which contains noisy \ac{GPS}
readings and the corresponding ground truth location of a moving Segway robot. For the filter process, we define the dynamic \ac{wna} model\cite{bar2004estimation}.
The only observable output is the
GPS position, thus the observation matrix is $\gvec{H}=\left(\begin{matrix}1 & 0\end{matrix}\right)$ and the observation noise covariance matrix is 
$\gvec{R}= {\gscal{r}^2}$.
%
%

The results reported in Fig.~\ref{fig:MSE noisy} examine the performance for synthetic data clean of outliers. In this case, it is shown that the \ac{oikf}-\ac{am} achieves the optimal minimal \ac{mse} bound, as most $\hat{\gamma}^2_t$ values are estimated to be zero, which means the model turns back to be the \ac{kf}. 
When the synthetic data is populated with outliers, the \ac{oikf}-\ac{am} has the best performance in terms of \ac{mse} compared to other algorithms for different values of observation noise variance $\gscal{r}^2$, as presented in Fig ~\ref{fig:MSE noisy with outliers}. More specifically, it coincides with the \ac{oikf}-\ac{em}, and even outperforms it for low observation noise, without using a second-order moment as in \ac{em}.

We proceed to the NCLT data set and demonstrate the tracking of a single trajectory in
Fig.~\ref{fig:position estimation}. We observe in Fig.~\ref{fig:position estimation}  the robustness of \ac{oikf}-\ac{am} in estimating the position from real-world data while reliably smoothing the outliers.
Table~\ref{tbl:NCLT results} presents the obtained position \ac{rmse} and \ac{mse} for each of the algorithms when the observation noise variance is set to $\gscal{r}^2=3^2$, which is a GPS "textbook" error, while $\gscal{r}^2$ is selected for each algorithm separately by grid search to yield the lowest \ac{mse}. For both scenarios $\gscal{q}^2$ was optimized by grid search.
We can see in Table~\ref{tbl:NCLT results} that our proposed \ac{oikf}-\ac{em} and \ac{oikf}-\ac{am} produce the lowest estimation errors for both settings, when the latter coincides with the \ac{oikf}-\ac{em}, even without using the second-order moment.
Furthermore, the OIKF-AM exhibits the shortest runtime in the domain of algorithms for outlier detection and weighting (except the $\chitest$ that only detects and then rejects the outliers). For instance, in comparison to our other suggested method the \ac{oikf}-\ac{em}, the \ac{oikf}-\ac{am} showcases an almost $40\%$ reduced runtime compared to it.

\vspace{-2mm}
\begin{table}[h]
\captionsetup{justification=centering,margin=0.3cm}
\caption{Position error for different values of $\gscal{r}^2$}
\vspace{-6mm}
\begin{center}
{\footnotesize
\setlength\tabcolsep{4.5pt}
\begin{tabular}{|m{1.25cm}|C{1.1cm}|C{1cm}|C{1.1cm}|C{1cm}|C{1cm}|}
\hline
& \multicolumn{2}{c|} {\small {\cellcolor{gray} $
{\gscal{r}^2=3^2}$}}  &   \multicolumn{2}{c|}{{\small\cellcolor{gray}optimal ${\gscal{r}^2}$ }}& {\cellcolor{lightgray}Runtime}  \\
&{\scriptsize {\cellcolor{lightgray}\ac{rmse}[m]}  } & {\scriptsize {\cellcolor{lightgray}\ac{mse}$\dB$} } &  {\scriptsize {\cellcolor{lightgray}\ac{rmse}[m]}} & {\scriptsize {\cellcolor{lightgray}\ac{mse}$\dB$ }}  & {\cellcolor{lightgray}[ms]}  \\
\hline
{\footnotesize  Noisy GPS} &  349.31   &  50.86  & 349.31  &  50.86  & -    \\
{\footnotesize \ac{kf}}   &  94.35    &  39.49  & 92.28   &  39.3 & 0.05 \\
{\footnotesize ORKF}      &  75.79    & 37.59   &  27.74     & 28.86 &  2.83  \\
$\chitest$ &    45.3  &    33.12  &    14   &   22.92    & 0.06  \\
{\footnotesize \ac{oikf}-\ac{am}}    &  \textbf{ 10.87} &     \textbf{20.72}  &    \textbf{10.38}      &   \textbf{20.33}    & \textbf{0.28}\\
{\footnotesize \ac{oikf}-\ac{em}}  &   \textbf{10.73}       &     \textbf{20.61} &    \textbf{10.35} &   \textbf{20.29} & 0.44   \\
\hline      
\end{tabular}  
\label{tbl:NCLT results}
}
\vspace{-9mm}
\end{center}
\end{table}
%
%
\section{Conclusions}\label{sec:Conclusions}
\vspace{-2.5mm}
In this work we derived a novel approach for outlier insensitive Kalman-filtering \acl{oikf}. Based on sparse Bayesian learning concepts, we modeled outliers as NUV with \ac{am} or \ac{em} approaches, resulting in a sparse outlier detection. 
Both algorithms are parameter-free and amount essentially to a short iterative process during the \emph{update step} of the \ac{kf}.
The presented empirical evaluations demonstrate that \ac{oikf}-\ac{am} and \ac{oikf}-\ac{em} present better performance compared to the other algorithms in terms of \ac{mse} and \ac{rmse}, highlighting the robustness and accuracy of \ac{oikf} for systems that rely on high-quality sensory data.
%
%
\bibliographystyle{IEEEtran}
\bibliography{IEEEabrv,OIKF}
\end{document}